\documentclass[prc,superscriptaddress,noshowpacs,unsortedaddress,twocolumn,showpacs,preprintnumbers,amsmath,amssymb]{revtex4-1}

\usepackage[pdftex]{graphicx,xcolor}
\usepackage{amsmath,amssymb,times}
\usepackage{color}

\begin{document}
\title{Wobbling Motion in the Many-Nucleon System}

\author{Masayuki~Matsuzaki}
\affiliation{Department of Physics, Fukuoka University of Education,
Munakata, Fukuoka 811-4192, Japan}

\date{\today}

\begin{abstract}
The purpose of this Perspective is not to present a new progress but to shed light on 
another stream of the study of the nuclear wobbling motion from a viewpoint of the many-body theory 
under an external field. Special attention is paid to the moments of inertia, which are microscopically 
calculated as a response of the full $A$-body system to rapid rotations described by the cranked RPA 
formalism. They depend on the rotation frequency, and it is shown that approach of two moments of inertia 
leads to the instability of rotating triaxial mean fields. 
\end{abstract}

\maketitle

\section{Introduction}
Exploring how matter responds to extreme environments has been always one of frontiers 
in physics and physical sciences. 
The atomic nucleus, which is governed by strong interaction, is also the case. 
In particular, when probed with low energies, the nucleus must be treated as a strongly 
correlated, finite many-body system. Mayer-Jensen's independent-particle picture 
in the selfconsistently formed mean field, the shell model, describes the ground and 
lowlying states in medium and heavy nuclei very well. The shape of the nucleus or that of 
the mean field is spherical at the magic numbers as those of electrons in the atom. 
When nucleon numbers deviate from the magic numbers in the shell, axially symmetric quadrupole 
deformation is expected. 
When large angular momenta are introduced to such a system, for example, by 
peripheral reactions with low incident energies, say, 10$A$ MeV, 
the nucleus collectively 
rotates about an axis perpendicular to the symmetry axis. This rotation is one-dimensional 
and uniform. As this rotation becomes rapid, the centrifugal force drives the shape to triaxial, 
and consequently, the nucleus can possess angular-momentum components about other axes. 
The resulting non-uniform rotation is the wobbling motion well known in classical 
mechanics \cite{landau}. 

The lowest-energy states with given angular momenta are called \textit{yrast} states. 
The spectra of such states experimentally define the moment of inertia $\cal{J}$ of the rotating body. 
Historically, there have been two models for nuclear matter in the 
rotating drop; the irrotational (irr) model and the rigid-body (rig) model (see Fig. 6A-2 
in Ref. \cite{Bohr1975Book}). 
Their absolute magnitudes can be related to the axially-symmetric deformation parameter 
$\beta$ but it has been known that 
$\cal{J}^\mathrm{(irr)} < \cal{J}^\mathrm{(exp)} < \cal{J}^\mathrm{(rig)}$, 
where $\cal{J}^\mathrm{(exp)}$ is determined from the energies of rotational 2$^{+}$ states in 
even-even nuclei. Their dependence on the deformation parameter can be extended to triaxial 
cases. 
When the nucleus stands still, the triaxial parameter $\gamma$ runs from 0$^\circ$ to 60$^\circ$, 
where 0$^\circ$ and 60$^\circ$ indicate prolate and oblate axially-symmetric shapes, respectively. 
When a rotation sets in, $\gamma$ needs three times wider range depending on the relation 
between the shape and the direction of rotation because triaxial nuclei can rotate about any 
axes, and therefore three moments of inertia (MOI) become relevant. 

\begin{figure}[htbp]
\begin{center}
\includegraphics[width=0.4\textwidth,clip]{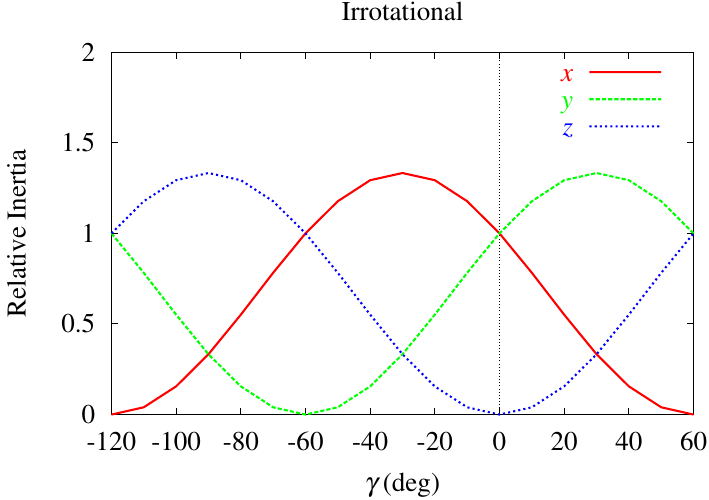}
 \caption{ Irrotational moments of inertia in the Lund convention normalized to $\cal{J}^\mathrm{(irr)}_\mathit{x}(\mathrm{\gamma=0})$. }
 \label{Fig1}
\end{center}
\end{figure}

Now, there are two conventions for the definition of the range of $\gamma$. One is the  Bohr-Mottelson (BM) 
convention, 
$$\mathcal{J}^\mathrm{(irr)}_\kappa\propto\beta^2\sin^2{(\gamma-\frac{2\pi}{3}\kappa)},$$ 
where $\kappa=$1, 2, 3 denote the $x$, $y$, $z$ axes and $\gamma$ runs from $-60^\circ$ to 120$^\circ$. 
The other is the so-called Lund convention, widely adopted in the Strutinsky calculations 
those days and also in the present article hereafter, 
$$\mathcal{J}^\mathrm{(irr)}_\kappa\propto\beta^2\sin^2{(\gamma+\frac{2\pi}{3}\kappa)},$$ 
where $\gamma$ runs from $-120^\circ$ to 60$^\circ$ (Fig. \ref{Fig1}). 

This convention-dependent terminology was explicitly used for describing physics as explained below. 

\section{Wobbling models without valence particle}
\subsection{Asymmetric Rotor Model}
By abstracting such underlying nucleon dynamics and specific forms of MOI, 
Bohr and Mottelson (Sect. 4-5e of Ref. \cite{Bohr1975Book}) discussed 
a small amplitude deviation from a one-dimensional rotation in terms of an asymmetric rotor 
with given MOI. It should be noted that non-uniform rotations were first investigated 
in the field of chemical physics \cite{Golden1949JCP} as clearly stated in Ref. \cite{Bohr1975Book}. 
The settings in Ref. \cite{Bohr1975Book} are briefly summarized as follows. 
All the necessary assumption is that the main rotation occurs about the axis of the largest inertia. 
Throughout this article, $\hbar$ is set to 1. 

\begin{itemize}
\item Three axes are numbered accordingly to MOI, in descending order such as 
$\cal{J}_\mathrm{1} > \cal{J}_\mathrm{2} > \cal{J}_\mathrm{3}$, 
then, $I_{1}\simeq I \gg 1$ for the yrast states, where $I$ is the total angular momentum. 

\item The commutation relation of the angular-momentum components in the principal-axis (PA) 
frame $[I_{-},I_{+}]=2I_1\simeq 2I$ holds, where $I_\pm=I_{2}\pm iI_{3}$. 

\item In terms of bosonic creation and annihilation operators, 
$c^\dagger=\frac{1}{\sqrt{2I}}I_{+}$, $c=\frac{1}{\sqrt{2I}}I_{-}$, $[c,c^\dagger]\simeq 1$, 
the Hamiltonian of the rotor 
$H=\sum_{j=1}^{3}\frac{1}{2\cal{J}_\mathit{j}}I_{j}^2$ 
can be cast into the form 
$\frac{1}{2\cal{J}_\mathrm{1}} I(I+1)+(\hat{n}+\frac{1}{2})\omega_\mathrm{wob}$
by diagonalizing the cross terms such as $c^\dagger c^\dagger$ and $c c$, 
where $\hat{n}$ is the number operator of the wobbling phonon and its eigenvalues 
are constrained as $(-1)^{n+I}=1$ (in even-even nuclei) by signature symmetry with respect to 
$\pi$ rotation about the first axis. 

\item The resulting frequency of the harmonic wobbling excitation is 
$\omega_\mathrm{wob}=\frac{I}{\cal{J}_\mathrm{1}}\sqrt{\frac{(\cal{J}_\mathrm{1}-\cal{J}_\mathrm{2})(\cal{J}_\mathrm{1}-\cal{J}_\mathrm{3})}{\cal{J}_\mathrm{2}\cal{J}_\mathrm{3}}}$, 
as in the classical mechanics. 

\end{itemize}
These are the bases of all the following discussions. 

Right after Bohr and Mottelson, Andersson et al. \cite{Andersson1976NPA} made an exploratory calculation 
of wobbling bands in $^{118}$Te at $I$ = 40 - 60  by adopting the rigid-body MOI. 
Its background was Lund group's calculations that predicted triaxial shapes triggered 
by rapid rotations. 
Liotta and Sorensen \cite{Liotta1978NPA} invoked the BM model above to study multiplicities 
and continuum gamma rays following heavy-ion reactions, paying attention to its fast 
electric quadrupole ($E$2) transitions, see Fig. 1 in Ref. \cite{Shimizu2004JCNP}. 
Theoretically, $\gamma$ vibrations around triaxial deformed shapes are possible. Based on 
this, it was proposed by Mikhailov and Janssen \cite{Mikhailov1978PLB72} that wobbling might be found as a 
high-spin continuation of the odd-spin sequence of the $\gamma$ band, just before the full 
cranked random-phase approximation (RPA) theory of Marshalek \cite{Marshalek1979NPA331}. This is consistent with 
that wobbling motion in the PA frame can be described as a shape fluctuation in the uniformly rotating (UR) 
frame as stated below, see Fig. 3 in Ref. \cite{Shimizu2004JCNP}. 

\subsection{Cranked RPA} 
The first step of the cranked RPA calculation is the UR mean field obtained by 
cranking about an axis, named $x$; 
$H'=\sum_{a}^{A}(h-\omega_\mathrm{rot}J_x)_a$, where $h$ is the Nilsson plus BCS Hamiltonian, for example. 
The commutation relation among the angular-momentum components is the same as that in the laboratory frame. 
The yrast configuration, which is the vacuum of the quasiparticle excitations, is implicitly regarded as even-even systems, 
but can be straightforwardly set to odd systems (see Eq. (4) in Ref. \cite{Matsuzaki2002PRC69}). 
The cranking term in the one-body Hamiltonian, 
$-\omega_\mathrm{rot}J_x$, acts on all the nucleons even in the cases with the vacuum is taken to be odd-$A$. 
This includes the centrifugal 
force causing alignments of individual nucleon's angular momenta to the rotation axis 
and leading to collective rotation. 
Residual two-body interactions among nucleons in the rotating mean field, $H_\mathrm{int}$, 
the doubly-stretched $Q$-$Q$ for example, 
are taken into account within the RPA, that is a linear response theory, 
in the form of the equation of motion, 
$\left[H'+H_\mathrm{int},X_n^\dagger\right]_\mathrm{RPA}=\omega_n X_n^\dagger$, 
for the eigenmode $X_{n}^\dagger$, which are superpositions of two-quasiparticle states. 
Importantly, that for the negative signature (odd spin in the even-even case) sector can be cast into the form 
of the implicit function, 
\begin{gather*}
(\omega_n^2-\omega_\mathrm{rot}^2)\left[\omega_n^2-\omega_\mathrm{rot}^2
\frac{\left(\mathcal{J}_x-\mathcal{J}_y(\omega_n)\right)
   \left(\mathcal{J}_x-\mathcal{J}_z(\omega_n)\right)}
{\mathcal{J}_y(\omega_n)\mathcal{J}_z(\omega_n)}
                                  \right] \notag \\
=0 ,
\end{gather*}
where the argument $(\omega_n)$ stresses that $\mathcal{J}_{y,z}$ are calculated mode by mode and the superscript 
$\mathrm{(eff)}$ in Ref. \cite{Matsuzaki2002PRC69} is omitted here. 
Their complete expressions in terms of two-quasiparticle energies and matrix elements and $\omega_n$ are given in 
Ref. \cite{Matsuzaki1990NPA509}. 

Its essence to describe wobbling motion is 
as follows \cite{Marshalek1979NPA331,Janssen1979NPA318,Shimizu1995NPA588}: 

\begin{itemize}
\item The RPA in the UR frame gives many individual (modified) two-quasiparticle- as well as 
collective vibrational normal modes and a Nambu-Goldstone (NG) mode with $\omega_{n=\mathrm{NG}}=\omega_\mathrm{rot}$ 
to restore rotational symmetry. 

\item The PA frame is defined by vanishing off-diagonal quadrupole moments. 
The two coordinate frames are related in terms of dynamical Euler angles. 
In the RPA order, this results in 
\begin{equation*}
\begin{split}
(J_x)_\mathrm{PA}&=(J_x)_\mathrm{UR} , \\
(iJ_y)_\mathrm{PA}&=(iJ_y-I\frac{Q_z}{2R^2\alpha_z})_\mathrm{UR} , \\
(J_z)_\mathrm{PA}&=(J_z+I\frac{Q_y}{2R^2\alpha_y})_\mathrm{UR} ,
\end{split}
\end{equation*}
where 
$Q_y=-c_0\sum_{a}^{A}(xz)_{a}$ and $Q_z=ic_0\sum_{a}^{A}(xy)_{a}$ 
are off-diagonal quadrupole operators, and 
$2R^2\alpha_y=c_0\langle x^2-z^2 \rangle$ and $2R^2\alpha_z=c_0\langle x^2-y^2 \rangle$ 
are static deformations, with $c_0=\sqrt{\frac{15}{4\pi}}$, see Ref. \cite{Shimizu1995NPA588} for details of the notations. 
It should be noted that the factors ``$\frac{1}{2}$" in both of Eq. (4.1) there are redundant. 

\item The transformation above recovers the commutation relation in the PA frame, 
$\langle [(iJ_y)_\mathrm{PA},(J_z)_\mathrm{PA}]\rangle=I=\langle (J_x)_\mathrm{PA} \rangle$. 

Further, $(J_{y,z})_\mathrm{PA}$ can be expanded in terms of RPA eigenmodes $X_{n}^\dagger$ and $X_{n}$ 
excluding the NG mode. This is the reason why the angular-momentum dynamics in the PA frame 
can be described in the RPA based on the UR mean field. It should be noted here that the 
condition of the vanishing off-diagonal quadrupole moments leads to the separation of the 
NG mode \cite{Marshalek1979NPA331,Shimizu1995NPA588,oni,Kaneko1992PRC}. 

\end{itemize}

\section{Wobbling in odd-$A$ nuclei}
\subsection{Experimental evidence}
Prerequisite for wobbling motion is equilibrium triaxial deformations by definition. 
The existence of such deformation has been a long dispute since 1950's \cite{Wilets1956PR,Davydov1958NP}. 
Theories such as the angular-momentum projection and the Strutinsky shell correction result 
in small but non-zero triaxial deformations, say, $\gamma=5^\circ$, in many cases. 
But it must be larger than the zero-point vibration in order that the picture of asymmetric 
rotor is meaningful, see Sect. 4.4 of Ref. \cite{Bohr1975Book}. 
In the so-called $\gamma$-soft nuclei, it is expected that triaxial shapes are stabilized by 
alignments. According to this expectation, the experimental data of $^{182}$Os and $^{124}$Xe 
after the $s$-band crossing were examined in Ref. \cite{Shimizu1995NPA588} but the results were not conclusive. 
And accordingly, no information was available about in what 
nuclei, at what spins, and with what $\gamma$, wobbling motion appears. 

Owing to the great developments of gamma ray detectors such as Euroball from the 1990's, the 
first firm data on wobbling excited on top of the triaxial super-/strongly deformed (TSD) 
band in $^{163}$Lu was reported in 2001 \cite{Odegard2001PRL86}. 
In Ref. \cite{Odegard2001PRL86}, the wobbling band, named TSD2, is excited on top of the yrast TSD band, TSD1, 
which is expected to be strongly deformed such as $\beta\sim 0.4$ and $\vert\gamma\vert\sim 20 ^\circ$. 
Note here that the sign of $\gamma$ is important as discussed below. 
In addition to the characteristic energy spectra; parallel to TSD1 and the opposite signature 
quantum number, the interband $E2$ transition probabilities, 
$B(E2: I_\mathrm{TSD2}\rightarrow (I-1)_\mathrm{TSD1})$ that prove large collectivity and to which 
Liotta and Sorensen paid attention, were also reported in the form of ratios to the inband 
rotational $B(E2: I\rightarrow I-2)$. Similar bands were also found in nearby nuclei, $^{161, 165, 167}$Lu and 
$^{167}$Ta \cite{Bringel2005EPJA,Shoenwasser2003PLB,Amro2003PLB,Hartley2009PRC}. 
These experimental results that wobbling excitations were found in odd-$A$ nuclei, rather than 
in even-even nuclei being expected to have simpler structure, opened up rich theoretical 
studies. 

\subsection{Theoretical models}
The first realistic calculation was done by Hamamoto \cite{Odegard2001PRL86,Hamamoto2002PRC65} 
in terms of the particle-rotor model (PRM), given by (1) replacing $I_{j}$ in the Hamiltonian 
of the asymmetric rotor model by $R_{j}=I_{j}-j_{j}$, where $R_{j}$,  $I_{j}$ and $j_{j}$ are the $j$-th components 
of the angular momentum of the rotor, the total system and the valence unpaired particle in the single-$j$ shell, 
respectively, 
and (2) adding its deformed potential. The former procedure produces the Coriolis coupling $\propto I_{j}j_{j}$. 
The essential input parameters in the PRM are MOI as in the original asymmetric rotor model. 
Those of the irrotational model, whose $\gamma$ dependence is accepted aside from its absolute 
magnitude, are often adopted. 

For the Lu isotopes in which wobbling excitations were observed, the equilibrium shape is 
understood to be +20$^\circ$ (in the Lund convention) because of the shape driving effect 
of the odd-$\pi_{13/2}$. 
This means $\cal{J}_\mathit{y} > \cal{J}_\mathit{x} > \cal{J}_\mathit{z}$ as can be seen in 
Fig. \ref{Fig1}, if the irrotational model is adopted. 
This would lead to an imaginary 
$\omega_\mathrm{wob}\propto\sqrt{(\cal{J}_\mathit{x}-\cal{J}_\mathit{y})(\cal{J}_\mathit{x}-\cal{J}_\mathit{z})}$. 
In order to avoid this, Hamamoto \cite{Hamamoto2002PRC65} inverted the sign of $\gamma$ 
in the irrotational MOI of the rotor. The result reproduced the characteristics of the data 
including the collective interband transitions well. But the physical reasoning of this 
inversion was not clear. 

Matsuzaki et al. \cite{Matsuzaki2002PRC65} found from the viewpoint of the nucleonic mean field that 
the physical reason why wobbling exists in odd-$A$ nuclei could be traced back to character of the models; 
only the rotor that represents the $A-1$ (even) nucleons rotates in the PRM, whereas full $A$ 
nucleons rotate in the cranked RPA description. They confirmed that 
$\cal{J}_\mathit{x}=\langle \mathit{J}_\mathit{x}\rangle /\omega_\mathrm{rot}$ 
of the full $A$-body could become larger than $\cal{J}_\mathit{y}$ by virtue of the alignment of 
the last odd proton in the high-$j$ orbital, in spite of the main rotation about the shortest axis in the 
positive-$\gamma$ rotation scheme. With increasing $\omega_\mathrm{rot}$, its contribution is getting 
relatively weak in comparison with that of increasing collective angular momentum. 
Then, $\cal{J}_\mathit{x}-\cal{J}_\mathit{y}$ $(>0)$ approaches zero, 
as in the cases of Fig. 5 in Ref. \cite{Shimizu2004JCNP}, consequently $\omega_\mathrm{wob}$ does. 
In other words, the wobbling excitation becomes unstable. 

This is contrary to a naive expectation 
that it is proportional to the total angular momentum $I$ provided MOI do not depend on $I$. 
This is a characteristic of the 
wobbling excited around the shortest-axis rotation. This mechanism can be translated into the PRM 
picture as shown in Ref. \cite{Shimizu2004JCNP}. 
Similar expressions were presented in Ref. \cite{Frauendorf2014PRC} ten years later. 

This work of Frauendorf and D\"{o}nau \cite{Frauendorf2014PRC} clarified the two physically different types 
of wobbling motion described in rather complicated convention-dependent terminology such as 
``wobbling on positive-$\gamma$ rotation" so far. As discussed above, 
$\cal{J}_\mathit{x} > \cal{J}_\mathit{y,z}$ must be fulfilled for wobbling excitation to exist when 
the main rotation is about the $x$ axis. In usual situations, collective rotation occurs about 
the axis of the intermediate length --- the negative-$\gamma$ rotation according to the Lund 
convention. The wobbling excitation in this scheme, which was considered by Bohr and Mottelson, 
was named longitudinal wobbling. On the other hand, collective rotation occurs about 
the shortest axis in the cases of the positive-$\gamma$ rotation according to the Lund convention. 
The situation $\cal{J}_\mathit{x} < \cal{J}_\mathit{y}$ that prevents wobbling from occurring is dissolved by the 
alignment of the last odd nucleon as shown in \cite{Matsuzaki2002PRC65}. 
The wobbling excitation in this scheme, which was thought to be realized in Lu isotopes, 
was named transverse wobbling. Consequently, two types of wobbling can be transparently 
designated with the shape parameter $\gamma$ runs only 0$^\circ$ - 60$^\circ$, by not restricting 
the main collective rotation to the $x$ axis. 

\section{Perspective}
Generally speaking, there are two paths to approach when one investigates an unknown phenomenon: 
One is to set up a phenomenological variable directly relevant to that phenomenon and the 
other is to look for how that phenomenon emerges from underlying degrees of freedom in the 
sub hierarchy. In the case of nuclear wobbling motion in odd-$A$ nuclei, the former corresponds 
to the PRM and the latter to the cranked RPA. 
The former is a fully quantum model for two vector degrees of freedom, but MOI, the most decisive 
internal quantities to determine experimental observables, energy spectra and transition probabilities 
between energy levels, are input parameters. 
Recently, fingerprints of wobbling motion in nuclei with normal or weak deformations in various mass regions are reported. 
Results of various input MOI for these data are intensively studied in Ref.~\cite{zhang2022}, for example. 

The latter is a quantum many-body model but the rotation is introduced as an external field. 
Whether collective wobbling excitation appears or not is determined as a response of the many-body system to the external field. 
When a wobbling state appears among huge number of normal modes, MOI are 
calculated at the same time as its excitation energy, as a linear response of the whole nucleus to the externally exerted rotation. 

Its another virtue as a many-body model is to exhibit a general notion of symmetry breaking and 
associated soft modes, ``the phase transition of the mean field" in other words, 
although they are in an approximated manner because of the finiteness, 
\begin{itemize}
\item a spherical mean field
becomes unstable as the quadrupole vibration excited on top 
of it softens, then an axially-symmetric mean field substitutes, 

\item this axially-symmetric mean field can become unstable 
as the $\gamma$ vibration softens, then a triaxially-deformed mean 
field substitutes. 
\end{itemize}
Similarly to these well-known cases, 

\begin{itemize}
\item the one-dimensionally rotating triaxial mean field, (irrespective of the axis of rotation,) 
may become unstable as the wobbling mode softens, then a two- or three-dimensionally tilted-axis 
rotating mean field \cite{fb} substitutes \cite{cuy,heiss,Matsuzaki2004PRC}, 

\item as its variant in weakly deformed cases, 
a mean field rotating about the shortest axis becomes unstable as the transverse wobbling mode 
softens with $\cal{J}_\mathit{x} - \cal{J}_\mathit{y}\rightarrow\mathrm{0}$ (with an adequate relabeling), via a tilted-axis rotation, 
then that rotating about the intermediate axis, which is a usual collective rotation 
and might support the longitudinal wobbling mode, substitutes, 
\end{itemize}
are relevant for the wobbling modes. 

On the other hand, some limitations and future problems exist: 

\noindent Conceptually, since the one-dimensional rotation is semiclassically introduced and fluctuations 
around it are treated within a small amplitude approximation, 
\begin{itemize}
\item the cranking picture is not very accurate for relatively low spin states in which 
angular-momentum vectors are inferred to wobble with rather large amplitudes, 
\item multi-phonon states are beyond the scope, although anharmonicities in the spectra, 
$\omega_\mathrm{wob}$, seen in the rotating frame \cite{Matsuzaki2004PRC} were found experimentally \cite{Jensen2002PRL89}. 
\end{itemize}
Numerically, 
\begin{itemize}
\item beyond the traditional RPA calculations with schematic separable (pairing plus 
quadrupole-quadrupole or its generalization \cite{Shoji2006IJMPE}) 
residual interactions, the finite amplitude method (FAM) \cite{Nakatsukasa2007PRC} calculations based on the 
Skyrme (and other types) energy density functionals are already available for collective 
vibrational states \cite{Hinohara2013PRC}, for example. Three MOI including effects of residual interactions 
are desirable for non-uniform rotational motions. 
\end{itemize}

Deep understanding of the responses of the nuclear many-body systems to rapid rotations, 
in particular those related to the stability of the collective wobbling modes in triaxial nuclei, 
is expected as soon as the limitations addressed here are resolved. 

\bibliography{92_merge}

\end{document}